\documentclass[12pt,aps,showpacs]{revtex4}

\usepackage{epsf}
\input{axodraw.sty}
\input{feynarts.sty}

\def\beq{\begin{equation}}
\def\eeq{\end{equation}}
\def\beqn{\begin{eqnarray}}
\def\eeqn{\end{eqnarray}}
\def\bea{\begin{eqnarray}}
\def\eea{\end{eqnarray}}
\def\be{\begin{equation}}
\def\ee{\end{equation}}

\begin{document}

\voffset 1.25cm

\title{ Chargino Pair
Production at Linear Collider and
Split Supersymmetry}
\author{Shou-hua Zhu}
\affiliation{ Institute of Theoretical Physics, School of Physics,
Peking University, Beijing 100871, P.R. China \\
Ottawa-Carleton Institute for Physics,  Department of Physics,
Carleton University, Ottawa, Canada K1S 5B6}

\date{\today}

\begin{abstract}

Recently N. Arkani-Hamed and S. Dimopoulos proposed a
supersymmetric model \cite{Arkani-Hamed:2004fb}, dubbed ``Split
Supersymmetry" in Ref. \cite{Giudice:2004tc},  which can remove
most of the unpleasant shortcomings of TeV Supersymmetry. In this
model all scalars except one finely tuned Higgs boson are ultra
heavy while the neutralino and chargino might remain light in
order to achieve gauge coupling unification and accord with the
dark matter density. In this paper, we investigated the impact of
this new model on chargino pair production at next generation
linear colliders. Our numerical results show that this process can
be used to probe sneutrino mass up to $10$ TeV. Therefore, precise
measurements of chargino pair production at the linear colliders
could distinguish Split Supersymmetry from TeV Supersymmetry.

\end{abstract}

\pacs{12.60.Jv, 14.80.Ly}

\maketitle

\newpage

\section{Introduction}

The standard model (SM) of high energy physics has been rigorously
tested by experiments, especially during the last decade of
beautiful experiments at CERN, Fermilab and SLAC
\cite{Altarelli:2004fq}. It is fair to say that no clear
indication of new physics has emerged from the data. Furthermore,
the unknown Higgs boson mass in the SM is predicted to be
$m_H=113^{+62}_{-42}$ GeV  and  $m_H \le 237$ GeV at 95\% c.l. by
global fits of the data \cite{Altarelli:2004fq}.

Despite the unprecedented success of the SM, we believe that the
SM is an incomplete theory and new physics beyond the SM will
eventually emerge at some  scale higher than that of electro-weak
symmetry breaking. The possible new physics will contribute
quadratically to the mass of the fundamental Higgs boson due to
its scalar nature. This phenomenon is the notorious fine-tuning or
naturalness problem. In order to deal with this issue, several
solutions have been proposed since 1970's of the last century: (1)
new symmetries are introduced to protect the mass from quadratical
radiative corrections. The popular choices are the fermion-boson
symmetry (Supersymmetry) \cite{SUSY} or the global symmetry in
little Higgs models \cite{LH}. We stress that in these two
approaches the Higgs boson is naturally light. (2) Extra
dimensions with low-scale gravity \cite{LS} have lowered the scale
of new physics. (3) The Higgs boson is the composite particle of
new fermions which are strongly interacting at the TeV scale. It
should be emphasized that although the new strong dynamics
(Technicolor, Topcolor etc.) \cite{HR} solves the fine-tuning
problem and is also consistent with the data, it is difficult to
obtain the preferred light Higgs boson mass without fine-tuning.
If a light Higgs boson is observed from experiments, solution (3)
will be disfavored. (4) The Higgs boson may be indeed finely tuned
\cite{Arkani-Hamed:2004fb} and naturalness is no longer the
guiding principle. This can be accomplished in the newly proposed
supersymmetrical model \cite{Arkani-Hamed:2004fb}. In this paper
we will investigate the impact from this novel idea on collider
phenomenology.

The idea of the finely tuned Higgs boson was recently proposed by
N. Arkani-Hamed and S. Dimopoulos \cite{Arkani-Hamed:2004fb} who
were motivated by the fine-tuning issue of the Cosmological
Constant, which has been set on solid ground as a consequence of
the Wilkinson Microwave Anisotropy Probe (WMAP) observations
\cite{WMAP}. The finely tuned supersymmetric model was also dubbed
``Split Supersymmetry" (SS) in Ref. \cite{Giudice:2004tc}. In such
an approach the naturalness principle was replaced by one of gauge
coupling unification and dark matter density. In the SS all the
scalars except one finely tuned Higgs boson are ultra heavy while
the neutralino and chargino might remain light in order to achieve
gauge coupling unification and accord with the dark matter
density. It should be emphasized that SS can remove most of the
unpleasant diseases of TeV supersymmetry, e.g. excessive flavor
and CP violation, fast dimension-5 proton decay etc.
\cite{Arkani-Hamed:2004fb,Giudice:2004tc}. Since SS emerged it has
already stimulated several investigations
\cite{Giudice:2004tc,PHENO}, which were mainly focused on
cosmological aspects and gauge coupling unification. Although
cosmological studies can provide solid evidence for new physics,
it says relatively little about the properties of new particles.
Therefore  studies about the SS impact on collider phenomenology
are necessary. In this paper we investigate the chargino pair
production at next generation linear colliders, especially whether
we can distinguish SS from TeV Supersymmetry with this process. In
order to discriminate SS from the usual TeV Supersymmetry, we need
to know the mass of scalars which is one of the most important
characteristics of SS.

 It is clear that the ultra heavy scalars in SS
 can hardly be directly produced at either hadron or
linear colliders. Therefore we should examine the virtual effects
of scalars in neutralino and chargino production. We stress that
the measurement of neutrino and chargino production at the LHC
could provide important hints for SS: One of the most important
production mechanism for neutralino and charginos at the LHC, i.e.
as decay products of scalar quarks, is no longer allowed. However
precise studies of chargino and neutralino production with
relatively rare events in SS seem challenging due to the large QCD
backgrounds at hadron colliders. In contrast linear colliders are
more promising, especially  chargino production which can
potentially be used  to reconstruct the chargino properties
\cite{CLC} as well as to determine the sneutrino mass.

\section{Formalism}

Due to R-parity conservation, supersymmetrical particles can only
be produced in pairs. In Fig. \ref{feyn} we show the Feynman
diagrams for chargino pair production at linear colliders.
 This process occurs at tree level via s-channel photon and Z, as
well as t-channel sneutrino exchanges. The transition amplitude
can be expressed as \cite{CLC} \bea T(e^+e^- \rightarrow
\tilde{\chi}_i^- \tilde{\chi}_j^+) =\frac{e^2}{s} Q_{\alpha \beta}
\left[ \bar v(e^+) \gamma_\mu P_\alpha u (e^-)\right] \left[ \bar
u(\tilde{\chi}_i^-) \gamma_\mu P_\beta v(\tilde{\chi}_j^+) \right]
\eea with chiralities $\alpha,\beta= L, R=(1\mp \gamma_5)/2$. Here
$i,j=1,2$ represent the indices of the chargino. In this paper we
are concerned about the sneutrino contributions, so we will only
explicitly present their expressions, and the others can be found
in Ref. \cite{CLC}. The sneutrino contributes to only $Q_{LR}$,
which can be written as \footnote{The CP phase $\Phi_\mu$ is set
to 0.} \bea Q_{LR}=\left\{ \begin{tabular}{ll}
 $1+\frac{D_Z}{s_W^2 c_W^2}
\left(s_W^2-\frac{1}{2}\right) \left(
s_W^2-\frac{3}{4}-\frac{1}{4} \cos(2 \phi_R) \right)
+\frac{D_{\tilde{\nu}}}{4 s_W^2} \left( 1+\cos(2 \phi_R)\right)
 $&  for $(11)\equiv \tilde{\chi}_1^- \tilde{\chi}_1^+$
 \\
$1+\frac{D_Z}{4 s_W^2 c_W^2} \left(s_W^2-\frac{1}{2}\right) \sin(2
\phi_R) +\frac{D_{\tilde{\nu}}}{4 s_W^2} \sin(2 \phi_R)
 $ & for $(12)\equiv\tilde{\chi}_1^- \tilde{\chi}_2^+$
\\
$1+\frac{D_Z}{s_W^2 c_W^2} \left(s_W^2-\frac{1}{2}\right) \left(
s_W^2-\frac{3}{4}+\frac{1}{4} \cos(2 \phi_R) \right)
+\frac{D_{\tilde{\nu}}}{4 s_W^2} \left( 1-\cos(2 \phi_R)\right)
 $ & for $(22)\equiv\tilde{\chi}_2^- \tilde{\chi}_2^+$,
 \end{tabular}
 \right.
 \label{qlr}
\eea where $D_Z=s/(s-m_Z^2)$,
$D_{\tilde{\nu}}=s/(t-m_{\tilde{\nu}}^2)$, $s_W=\sin(\theta_W)$
and $c_W=\cos(\theta_W)$ with $\theta_W$ is the weak angle, and
\bea \cos(2 \theta_R)&=&-\frac{M_2^2-|\mu|^2+2 m_W^2
\cos(2\beta)}{\Delta_C},
\nonumber \\
\sin(2 \theta_R)&=&-\frac{2 m_W
\sqrt{M_2^2+|\mu|^2-(M_2^2-|\mu|^2)\cos(2 \beta)+2 M_2 |\mu|\sin(2
\beta)}}{\Delta_C} \eea with \bea \Delta_C &=&
\sqrt{(M_2^2-|\mu|^2)^2+4 m_W^4\cos^2(2 \beta)+ 4 m_W^2
(M_2^2+|\mu|^2)+8 m_W^2 M_2 |\mu| \sin(2 \beta)}. \eea Here $M_2$
and $\mu$ are $SU(2)$ gaugino  and higgsino mass parameters, and
as usual $\tan\beta=v_2/v_1$ is the ratio of the vacuum
expectation values of the two neutral Higgs fields which break
electroweak symmetry.

The unpolarized differential cross section  and left-right
asymmetry can be derived \cite{CLC} as \bea
\frac{d\sigma}{d\cos\Theta} &=& \frac{\pi \alpha^2}{2 s}
\lambda^{1/2} N \label{diff}
 \\
A_{LR}(\Theta) &=& \frac{N^\prime}{N}, \label{difflr} \eea where
$\Theta$ is the angle between the $e^-$ and $\tilde{\chi}_i^-$ and
\bea N &=& \left( 1-(\mu_i^2-\mu_j^2)^2+\lambda \cos^2\Theta
\right) Q_1 +4 \mu_i \mu_j Q_2 +2 \lambda^{1/2} \cos\Theta Q_3
\nonumber \\
N^\prime &=& N[Q_1 \rightarrow Q_1^\prime, Q_2 \rightarrow Q_2^\prime,
Q_3 \rightarrow Q_3^\prime]
\nonumber \\
\mu_i &=& \frac{m_{\tilde{\chi}_i}}{\sqrt{s}}, \ \
\lambda=\left(1-(\mu_i+\mu_j)^2\right) \left(1-(\mu_i-\mu_j)^2\right).
\label{NN}
\eea
Here
\bea
Q_1 &=& \frac{1}{4} \left[
|Q_{RR}|^2+|Q_{LL}|^2+|Q_{RL}|^2+|Q_{LR}|^2 \right]
\nonumber \\
Q_2 &=&  \frac{1}{2} {\rm Re}\left[ Q_{RR} Q_{RL}^*+
Q_{LL} Q_{LR}^* \right]
\nonumber \\
Q_3 &=& \frac{1}{4} \left[
|Q_{RR}|^2+|Q_{LL}|^2-|Q_{RL}|^2-|Q_{LR}|^2 \right]
\nonumber \\
Q_1^\prime &=& \frac{1}{4} \left[
|Q_{RR}|^2-|Q_{LL}|^2+|Q_{RL}|^2-|Q_{LR}|^2 \right]
\nonumber \\
Q_2^\prime &=&  \frac{1}{2} {\rm Re}\left[ Q_{RR} Q_{RL}^*-
Q_{LL} Q_{LR}^* \right]
\nonumber \\
Q_3^\prime &=& \frac{1}{4} \left[
|Q_{RR}|^2-|Q_{LL}|^2-|Q_{RL}|^2+|Q_{LR}|^2 \right]. \label{qdef}
\eea From Eq. \ref{qdef} we can see that the sneutrino can
contribute to all the quartic charges $Q_i$ and $Q_i^\prime$
($i=1-3$).

The differential cross section $\frac{d\sigma}{d\cos\Theta}$ and
left-right asymmetry $A_{LR}(\Theta)$ are useful for large events
analysis, while for fewer events we can define the integrated
observables: the total cross section, forward-backward asymmetry
and integrated left-right asymmetry respectively. Based on Eq.
\ref{diff} and \ref{NN} they are
 \bea
 \sigma &=& \frac{\pi \alpha^2}{s}  \lambda^{1/2}
 \left(\left( 1-(\mu_i^2-\mu_j^2)^2\right) Q_1+ 4 \mu_i \mu_j
Q_2+\frac{1}{3} \lambda Q_1\right), \nonumber \\
A_{FB} &\equiv& \frac{\int_{-1}^0 \frac{d\sigma}{d\cos\Theta} d
\cos\Theta -\int_{0}^1 \frac{d\sigma}{d\cos\Theta} d
\cos\Theta}{\int_{-1}^0 \frac{d\sigma}{d\cos\Theta} d \cos\Theta
+\int_{0}^1 \frac{d\sigma}{d\cos\Theta} d \cos\Theta}=\frac{-
\lambda^{1/2} Q_3}{ \left( 1-(\mu_i^2-\mu_j^2)^2\right) Q_1+ 4
\mu_i \mu_j Q_2+\frac{1}{3} \lambda Q_1}
 \nonumber \\
A_{LR} &\equiv& \frac{\int_{-1}^1 N^\prime d
\cos\Theta}{\int_{-1}^1 N d \cos\Theta}= \frac{\left(
1-(\mu_i^2-\mu_j^2)^2\right) Q_1^\prime+ 4 \mu_i \mu_j
Q_2^\prime+\frac{1}{3} \lambda Q_1^\prime}{ \left(
1-(\mu_i^2-\mu_j^2)^2\right) Q_1+ 4 \mu_i \mu_j Q_2+\frac{1}{3}
\lambda Q_1}. \eea

\section{Numerical analysis}

In the following, we  study numerically the SS effects on the
observables of chargino pair production at linear colliders. The
SS parameter space will be analyzed first. Based on the gaugino
mass unification assumption and the dark matter density constraint
$0.094 <\Omega_{DM} h^2 <0.129$, the allowed parameter space of SS
can be classified into three categories \cite{Giudice:2004tc}
\begin{enumerate}
\item
Mixed region: $M_2$ and $\mu$ are comparable and below 1-2 TeV. In
this region the lightest neutralino and chargino are always in a
the mixed state of the higgsino and gaugino, and the mass of the
chargino and neutralino can be as low as the present experimental
limit. We stress that in this case the long-lived gluino might be
found at the LHC and gluino lifetime can be used to probe scale of
the ultra heavy scalar.

\item
$\mu \approx 1.0-1.2$ TeV and the gaugino masses are arbitrarily
larger than $\mu$. The neutralino and chargino are considerably
heavier.

\item
$M_2 \approx 2.0-2.5$ and the other parameters are arbitrarily
larger.

\end{enumerate}
It is clear that the case (2) and (3) are not happy for colliders,
so we will investigate the collider signature for case (1) only.
We stress that the discovery reach of the LHC and next linear
colliders can not cover the entire SS parameter space.

In the mixed region, the magnitude of both $M_2$ and $\mu$ can be
as low as $200$ GeV \footnote{See fig. 11 in Ref.
\cite{Giudice:2004tc}.}. We take two typical points labelled as
``{\bf Pa}" and ``{\bf Pb}" \bea {\rm \bf Pa}: && M_2=200\ {\rm
GeV}, \mu=200\ {\rm GeV}, \tan\beta=20,
\sin(2 \Phi_R)=-0.9655, \nonumber \\
&& \cos(2 \Phi_R)=0.2605, m_{\tilde{\chi}_1^\pm}= 147.5\ {\rm
GeV}, m_{\tilde{\chi}_2^\pm}= 266.8\ {\rm GeV};
\label{Pa} \\
{\rm \bf Pb}: && M_2=1000\ {\rm GeV}, \mu=600\ {\rm GeV},
\tan\beta=20,
\sin(2 \Phi_R)=-0.3497, \nonumber \\
&& \cos(2 \Phi_R)=-0.9369, m_{\tilde{\chi}_1^\pm}= 593.1\ {\rm
GeV}, m_{\tilde{\chi}_2^\pm}= 1010.5\ {\rm GeV}. \label{Pb}
 \eea The
characteristics for {\bf Pa} are the light chargino and
$|\sin(2\Phi_R)| \sim 1$ while for {\bf Pb} chargino is heavy and
$|\cos(2\Phi_R)|\sim 1$. It should be noted that the sign of $\mu$
does not affect chargino production.

In Fig. \ref{siga} we show the cross sections as a function of
sneutrino mass for {\bf Pa} with $\sqrt{s}=800$ GeV. It is obvious
that the three modes (11), (12) and (22) can all act as probes of
the sneutrino mass. This feature is typical in the mixed region
for low $M_2$. Moreover one can distinguish the 1 TeV and 10 TeV
sneutrino through the cross section measurements. For example, the
cross sections for the (11) mode are 0.18 pb and 0.25 pb for a 1
TeV and 10 TeV sneutrino respectively, and for the (22) mode they
are 0.19 pb and 0.23 pb respectively. We can also see that the
cross sections for the (11) and (22) modes increase while for the
(12) mode it decreases with  sneutrino mass. This is because as
the sneutrino mass increases, the contributions arising from
t-channel sneutrino decrease, as noticed in Ref. \cite{CLC}, and
the strong destructive effects for a light sneutrino in the (11)
and (22) modes are weaker than for a heavy sneutrino.

In Fig. \ref{afba} and \ref{alra} the differential cross sections,
$A_{FB}$, $A_{LR}(\Theta)$ and $A_{LR}$ are given for {\bf Pa}
respectively. From diagrams (A)-(C) of both figures we can see
that as the sneutrino mass increases, the differential cross
sections and left-right asymmetry tend generally to be more
symmetric, and again the three modes can all act as  probes of the
sneutrino mass. For $A_{FB}$, the (11) and (12) modes are more
sensitive to the sneutrino mass than that of the (22), while for
$A_{LR}$ the (12) mode is more sensitive.

In Fig. \ref{sigb} and \ref{b} we show the cross sections,
$A_{FB}$ and $A_{LR}$ as a function of sneutrino mass for {\bf Pb}
with $\sqrt{s}=3$ TeV. For the cross section and $A_{FB}$, the
(12) and (22) modes are sensitive to the sneutrino mass while the
(11) mode is not. Moreover the (22) mode is more promising than
that of the (12) because of its relatively large cross section.
For $A_{LR}$, only the (12) mode is sensitive to the sneutrino
mass. It is obvious that for all observables the (11) mode is not
sensitive to the sneutrino mass because, from Eq. \ref{qlr},
$\cos(2 \Phi_R) \sim -1$ makes sneutrino almost decouple from the
(11) mode.

\section{Discussions and conclusions}

We have investigated the impact of Split Supersymmetry on chargino
pair production at the next linear colliders. From the numerical
discussions for two typical parameter sets {\bf Pa} and {\bf Pb}
which are inferred from gauge coupling unification and constrained
by the dark matter density, we can see that chargino pair
production is sensitive to the sneutrino mass up to 10 TeV. For
low $M_2$ all three modes (11), (12) and (22) are sensitive to the
sneutrino mass, while for high $M_2$ the (22) mode is most
promising. We stress that the analysis of all three modes (11),
(12) and (22) is necessary in order to distinguish SS from TeV
Supersymmetry.

In this study, we have only included tree-level contributions and
are aware that higher order corrections \cite{Diaz:1997kv} are
likely to be non-negligible, especially for precise determination
on scalar mass scale. Nevertheless, we feel that our approach is
satisfactory for a preliminary study to gauge the potential of
this process for distinguishing SS from TeV Supersymmetry.

It is known that beam polarization will be a very useful tool at
the linear colliders to enhance signal, reduce background as well
as extract couplings etc. \cite{Moortgat-Pick:2004gf}. In this
paper we are more concerned about physical observable's dependence
on the sneutrino contributions which are only explicitly present
in $Q_{LR}$ in Eq. \ref{qlr}. It is obvious that the beam
polarization will enhance such kind of dependence, especially for
double beams polarization.

If SS stands up to the scrutiny of LHC (no light scalars, except
one Higgs boson, are found), TeV Supersymmetry will be at least
highly disfavored. In order to test SS, it is necessary to exam
the SUSY couplings relations at linear colliders, for example
$M_W^\chi=M_W$ in mixed region  \cite{Tsukamoto:1993gt}. It is
very promising that precise test on such kinds of SUSY relations
can be achieved \cite{Tsukamoto:1993gt}. We stress that the more
solid conclusions can be drawn only after the real simulations.


\noindent {\em Acknowledgements}: The author thanks S. Godfrey and
A. Fan for reading the manuscript carefully, and M. Peskin for
drawing my attention to Ref. \cite{Tsukamoto:1993gt}. This work
was supported in part by the Natural Sciences and Engineering
Research Council of Canada, as well as Natural Science Foundation
of China.

\begin{figure}
\unitlength=1bp%
\begin{feynartspicture}(432,504)(4,5.3)

\FADiagram{(1)} \FAProp(0.,15.)(6.,10.)(0.,){/Straight}{1}
\FALabel(2.48771,11.7893)[tr]{$e$}
\FAProp(0.,5.)(6.,10.)(0.,){/Straight}{-1}
\FALabel(3.51229,6.78926)[tl]{$e$}
\FAProp(20.,15.)(14.,10.)(0.,){/Straight}{1}
\FALabel(16.4877,13.2107)[br]{$\tilde \chi_j^+$}
\FAProp(20.,5.)(14.,10.)(0.,){/Straight}{-1}
\FALabel(17.5123,8.21074)[bl]{$\tilde \chi_i^-$}
\FAProp(6.,10.)(14.,10.)(0.,){/Sine}{0}
\FALabel(10.,8.93)[t]{$\gamma$} \FAVert(6.,10.){0}
\FAVert(14.,10.){0}

\FADiagram{(2)} \FAProp(0.,15.)(6.,10.)(0.,){/Straight}{1}
\FALabel(2.48771,11.7893)[tr]{$e$}
\FAProp(0.,5.)(6.,10.)(0.,){/Straight}{-1}
\FALabel(3.51229,6.78926)[tl]{$e$}
\FAProp(20.,15.)(14.,10.)(0.,){/Straight}{1}
\FALabel(16.4877,13.2107)[br]{$\tilde \chi_j^+$}
\FAProp(20.,5.)(14.,10.)(0.,){/Straight}{-1}
\FALabel(17.5123,8.21074)[bl]{$\tilde \chi_i^-$}
\FAProp(6.,10.)(14.,10.)(0.,){/Sine}{0} \FALabel(10.,8.93)[t]{$Z$}
\FAVert(6.,10.){0} \FAVert(14.,10.){0}

\FADiagram{(3)} \FAProp(0.,15.)(10.,14.)(0.,){/Straight}{1}
\FALabel(4.84577,13.4377)[t]{$e$}
\FAProp(0.,5.)(10.,6.)(0.,){/Straight}{-1}
\FALabel(5.15423,4.43769)[t]{$e$}
\FAProp(20.,15.)(10.,6.)(0.,){/Straight}{1}
\FALabel(16.8128,13.2058)[br]{$\tilde \chi_j^+$}
\FAProp(20.,5.)(10.,14.)(0.,){/Straight}{-1}
\FALabel(17.6872,8.20582)[bl]{$\tilde \chi_i^-$}
\FAProp(10.,14.)(10.,6.)(0.,){/ScalarDash}{1}
\FALabel(9.03,10.)[r]{$\tilde \nu_e$} \FAVert(10.,14.){0}
\FAVert(10.,6.){0}

\end{feynartspicture}
\caption[]{ Feynman diagrams for $e^+ e^- \rightarrow \tilde
\chi_i^- \tilde \chi_i^+$. } \label{feyn}
\end{figure}

\begin{figure}[thb]
\epsfxsize=12 cm \centerline{\epsffile{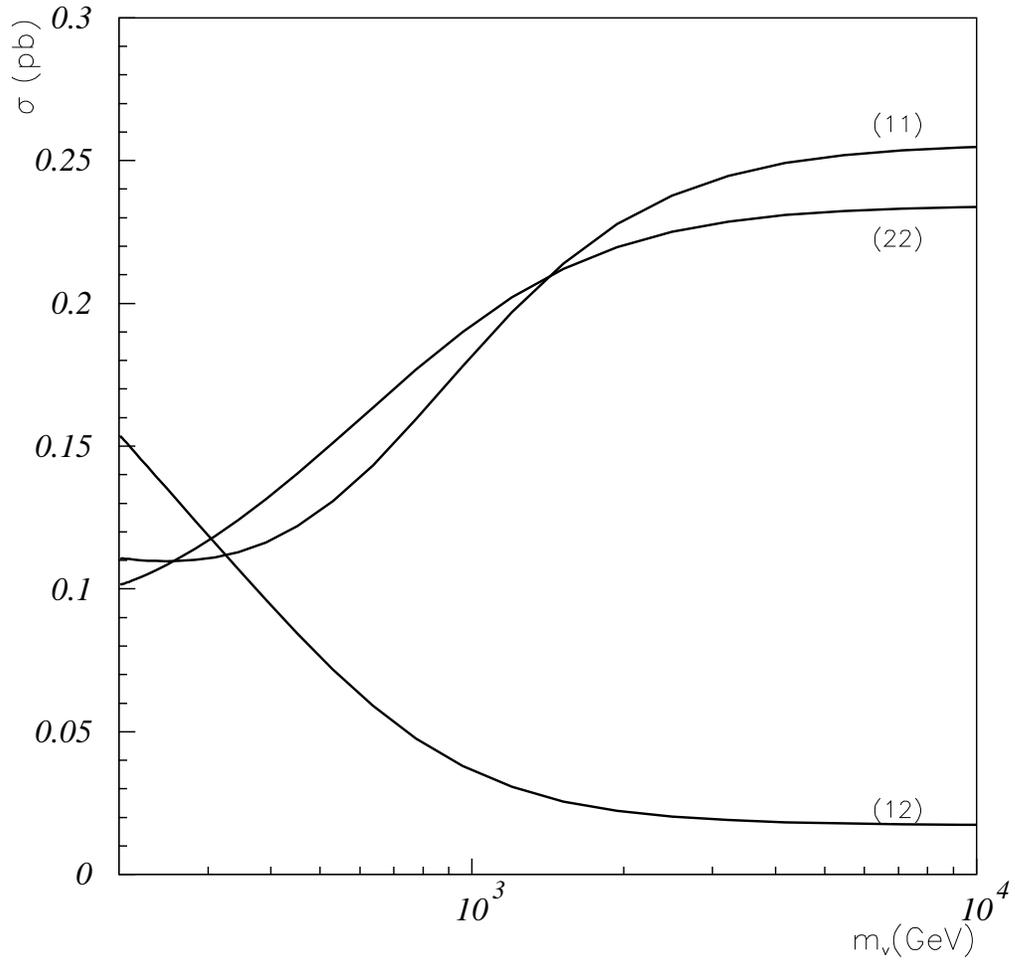}} \caption{Cross
sections as a function of sneutrino mass for {\bf Pa} with
$\sqrt{s}=800$ GeV. Definitions of (11), (12) and (22) are in Eq.
\ref{qlr}. } \label{siga}
\end{figure}

\begin{figure}[thb]
\epsfxsize=12 cm \centerline{\epsffile{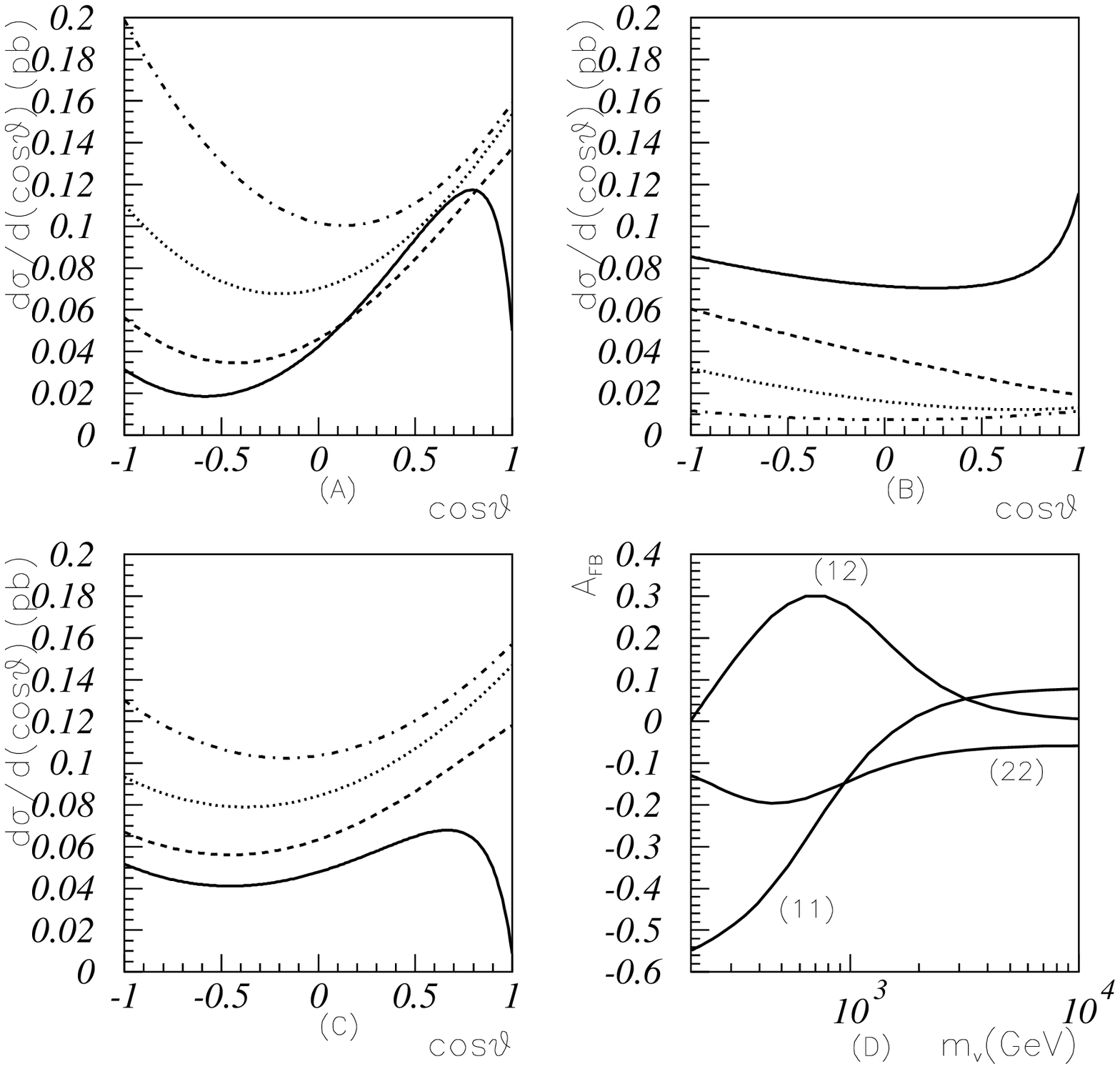}} \caption{
Differential cross sections [ (A)-(C)] and $A_{FB}$ [ (D) ] as
functions of $\cos(\Theta)$  and sneutrino mass for {\bf Pa} with
$\sqrt{s}=800$ GeV. For (A)-(C) solid, dashed, dotted and
dot-dashed lines represent $m_{\tilde{v}}=200, 500, 10^3, 10^4$
GeV respectively. } \label{afba}
\end{figure}

\begin{figure}[thb]
\epsfxsize=12 cm \centerline{\epsffile{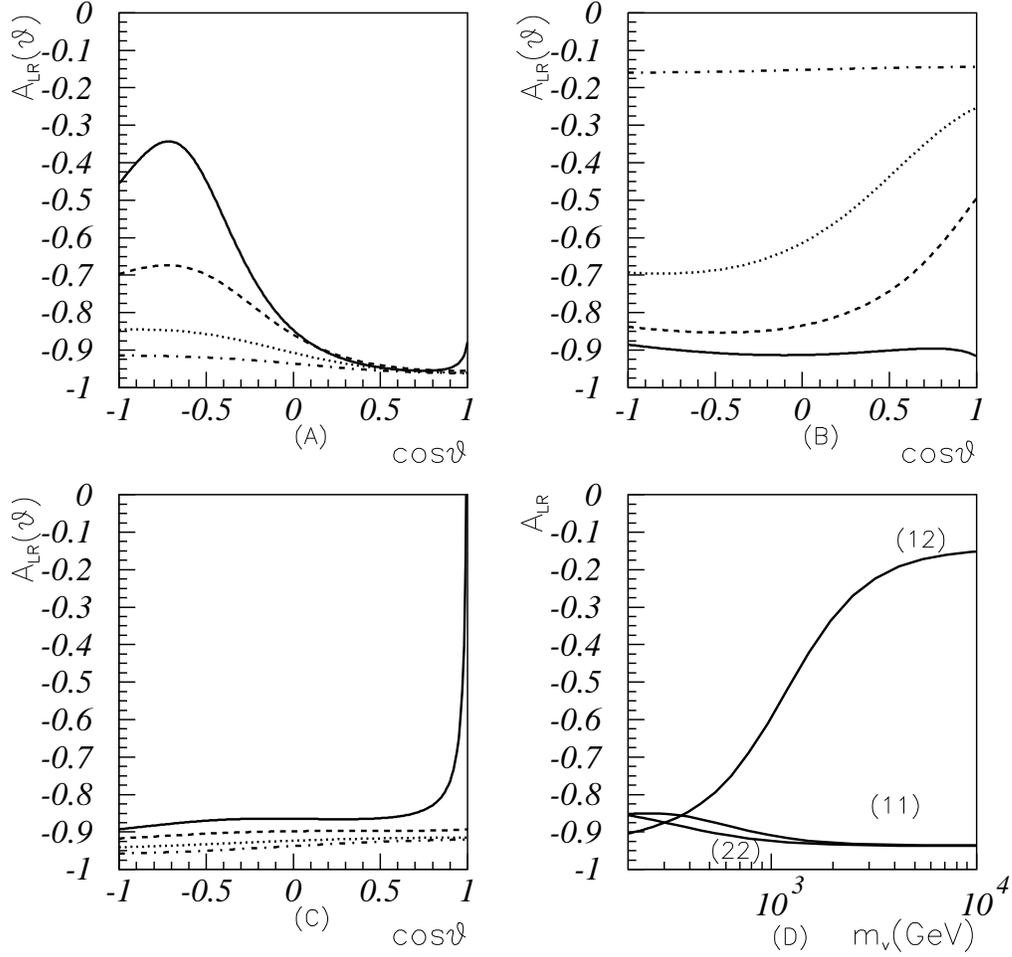}} \caption{
$A_{LR}(\Theta)$ [ (A)-(C)] and $A_{LR}$ [ (D) ] as functions of
$\cos(\Theta)$ and sneutrino mass for {\bf Pa} with $\sqrt{s}=800$
GeV. For (A)-(C) solid, dashed, dotted and dot-dashed lines
represent $m_{\tilde{v}}=200, 500, 10^3, 10^4$ GeV respectively. }
\label{alra}
\end{figure}

\begin{figure}[thb]
\epsfxsize=12 cm \centerline{\epsffile{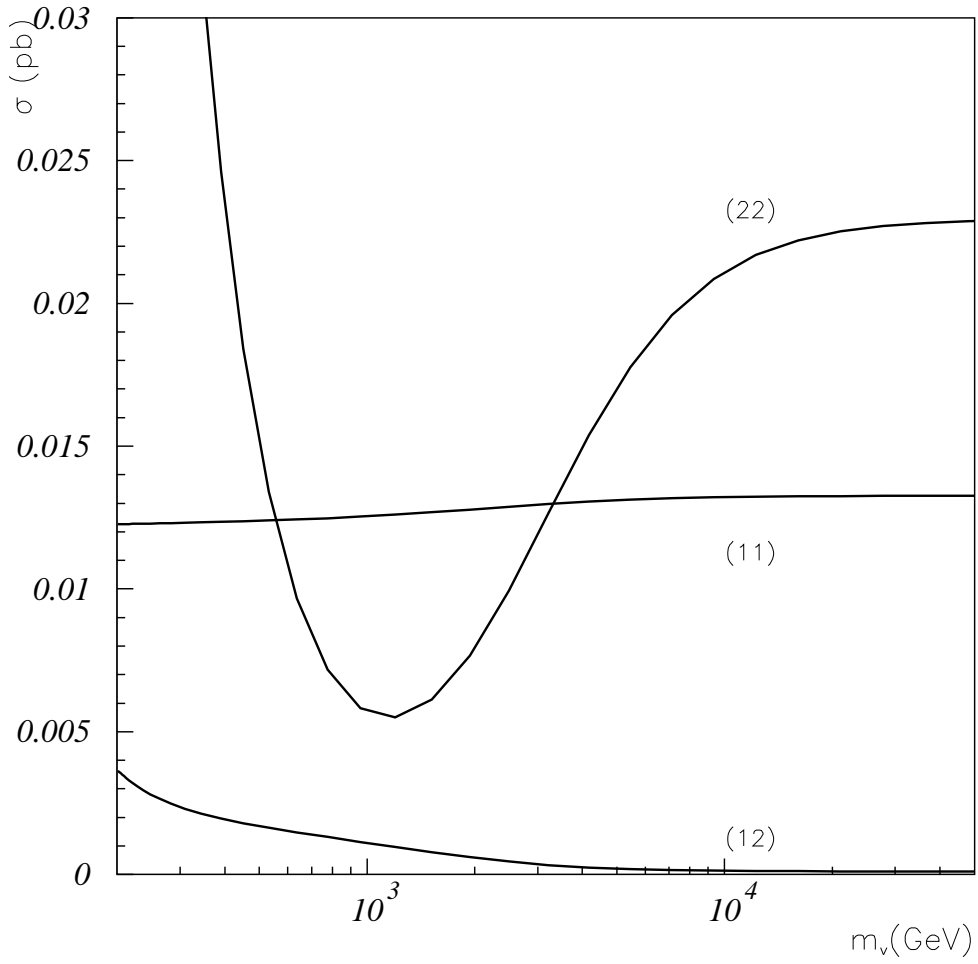}} \caption{Cross
sections as a function of sneutrino mass for {\bf Pb} with
$\sqrt{s}=3$ TeV.  } \label{sigb}
\end{figure}

\begin{figure}[thb]
\epsfxsize=12 cm \centerline{\epsffile{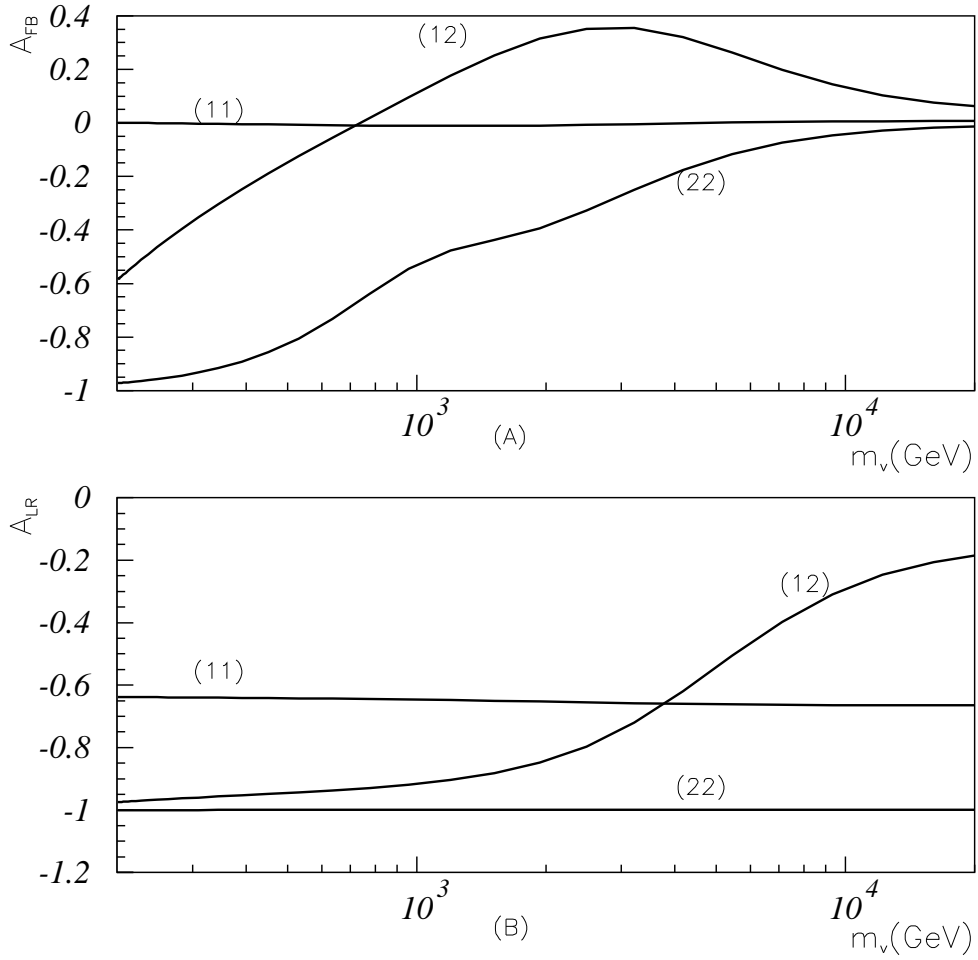}} \caption{ $A_{FB}$
[ (A) ] and $A_{LR}$ [ (B) ] as a function of sneutrino mass for
{\bf Pb} with $\sqrt{s}=3$ TeV.} \label{b}
\end{figure}

\end{document}